\documentclass{article}
\usepackage{amsmath,amsthm}
\usepackage{amssymb,latexsym}
\usepackage[mathscr]{eucal}
\usepackage{setspace}
\usepackage{graphics}
\usepackage{array}

\newcommand{\pa}{\partial}
\newcommand{\pr}{\prime}
\newcommand{\lp}{\left(}
\newcommand{\rp}{\right)}
\newcommand{\lb}{\left[}
\newcommand{\rb}{\right]}

\newcommand{\rc}{\right\}}
\newcommand{\la}{\left<}
\newcommand{\ra}{\right>}
\newcommand{\be}{\begin{equation}}
\newcommand{\ee}{\end{equation}}

\newcommand{\ihat}{\bf\hat{i}}
\newcommand{\nhat}{\bf\hat{n}}
\newcommand{\that}{\bf\hat{t}}
\newcommand{\bhat}{\bf\hat{b}}

\begin{document}

\title{\textbf{Vortex Motion in Superfluid $^4$He: Effects of Normal Fluid Flow}}         
\author{Bhimsen K. Shivamoggi\footnote{Permanent Address: University of Central Florida, Orlando, FL 32816-1364, USA; E-mail: bhimsen.shivamoggi@ucf.edu}\\
J. M. Burgers Centre and Fluid Dynamics Laboratory\\
Department of Physics\\
Eindhoven University of Technology\\
NL-5600MB Eindhoven, The Netherlands
}        
\date{}          
\maketitle

\large{\bf Abstract}

The motion of a vortex filament in superfluid $^4$He is considered by using the Hall-Vinen-Bekarevich-Khalatnikov (HVBK) (\cite{Hal}, \cite{Hal2} and \cite{Bek}) phenomenological model for the scattering process between the vortex and thermal excitations in liquid $^4$He. The HVBK equations are analytically formulated first in the intrinsic geometric parameter space to obtain insights into the physical implications of the friction terms, associated with the friction coefficients $\alpha$ and $\alpha^\pr$ (in the Hall-Vinen notation) as well as the previous neglect of the friction term associated with the friction coefficient $\alpha^\pr$. The normal fluid velocity components both {\it along} and {\it transverse} to the vortex filament are included. This analytical development also serves to highlight the difficulties arising in making further progress on this route. A reformulation of the HVBK equation in the extrinsic vortex filament coordinate space is then given which is known (Shivamoggi \cite{Shi}) to provide a useful alternative analytical approach in this regard. A nonlinear Schrodinger equation for the propagation of nonlinear Kelvin waves on a vortex filament in a superfluid is given taking into account the generalized normal fluid flow. The friction term associated with $\alpha^\pr$, even in the presence of the normal fluid velocity components {\it transverse} to the vortex filament, is shown to produce merely an algebraic growth of the Kelvin waves hence providing further justification for the neglect of this term. On the other hand, the instability produced by the friction term associated with $\alpha$ via the normal fluid velocity component {\it along} the vortex filament is shown to manifest itself as a parametric amplification on considering the problem of a rotating planar vortex filament in a superfluid.

\pagebreak

\noindent\Large\textbf{1. Introduction}\\

\large In the standard model of superfluid $^4$He (due to Landau \cite{Lan}), one considers the superfluid below the lambda point as an inviscid, irrotational fluid with thermal excitations moving upon that underlying fluid. These excitations are taken to constitute the normal fluid which interacts with the superfluid via mutual friction only in the presence of vortices\footnote{Vorticity in superfluid $^4$He is confined to vortices which, as Onsager \cite{Ons} suggested, are linear topological defects with the superfluid density vanishing at the vortex core. This allows liquid $^4$He to have the angular momentum as in a solid body rotation situation while maintaining irrotationality, as per Landau's \cite{Lan} theory, over almost the whole volume. This, on the other hand, as Onsager \cite{Ons} pointed out, leads to the result that the circulation around a vortex line is quantized, which was confirmed experimentally by Vinen \cite{Vin}. Rayfield and Reif \cite{Ray} also gave direct evidence for the existence of quantized vortices in superfluid $^4$He via charge carrying vortex rings having one quantum of circulation. Direct observation of quantized vortices has been accomplished in superfuidity manifested in dilute Bose-Einstein condensates produced by laser cooling in laboratory experiments (Pethick and Smith \cite{Pet}). Direct observation of vortex cores has also been accomplished recently (Bewley et al. \cite{Bew}) by using small solid hydrogen particles as traces in liquid $^4$He.}. The thermal excitations are scattered by the vortices when there is a relative velocity between them, hence giving rise to the so-called ``mutual friction" (Feynman \cite{Fey}). The mutual friction\footnote{Our understanding of the microscopic origin of the physical processes underlying the mutual friction is not good (Donnelly \cite{Don}) so we do not have an adequate theory of the fundamental roton-vortex scattering process in superfluid $^4$He yet.} was confirmed experimentally via the attenuation of second sound in uniformly rotating liquid $^4$He (Hall and Vinen \cite{Hal}, \cite{Hal2}). Upon making the plausible assumption that a thermal excitation can exchange momentum only in a direction perpendicular to the scattering vortex line, these experiments indicated (Vinen \cite{Vin2}) that the friction force will have a large constant value when the relative velocity is perpendicular to the rotation axis (with which the vortex lines are aligned) but will be zero when the relative velocity is parallel to the latter. Thanks to the thin cores\footnote{The core radius is of the order of quantum coherence length ($1^{\text o}$ Angstrom).} of vortices in liquid $^4$He, the detailed core physics does not seem to contribute to the long range dynamical effects of the vortex. As a result, vortices in superfluid $^4$He essentially behave like classical vortex filaments, barring the quantum mechanical features associated with their circulations and core radii\footnote{This scenario is violated in vortex reconnection events which involve sharp distortions of vortex lines (Paoletti et al. \cite{Pao}) and the concomitant generation of Kelvin waves associated with helical displacements of the vortex cores (Svistunov \cite{Svi}).}. The numerical simulations of Schwarz \cite{Sch} and \cite{Sch2}, which provided considerable insight into superfluid vortex dynamics, are based on the idea that, except on very short length scales, quantized vortices can be regarded as vortex filaments moving according to classical fluid dynamics, with the inclusion of the mutual friction force.

The dominant term in the vortex self-advection velocity according to Biot-Savart law in hydrodynamics is given by the local induction approximation (LIA) (Da Rios \cite{DaR}, Arms and Hama \cite{Arm}) in which the singularity due to the neglect of the finite vortex core size is resolved by an asymptotic calculation. Using the LIA, Da Rios \cite{DaR} and Betchov \cite{Bet} independently derived time evolution equations governing the inextensional motion of a vortex filament in an irrotational fluid in terms of its intrinsic geometric parameters - curvature and torsion. The Da Rios-Betchov equations were shown (Hasimoto \cite{Has}) to be combined to give a nonlinear Schrodinger equation. This equation admits a single-soliton solution (Zakharov and Shabat \cite{Zak}) which describes an isolated loop of helical twisting motion along the vortex line. The LIA is based on the assumption that the motion of the vortex filament is governed solely by the local features on the filament, so distant parts of the filament can not come close to each other during the motion. However, the large-amplitude solutions in LIA do not comply with this premise as in a self-interaction of the vortex filament during a vortex reconnection process where the non-local effects dominate\footnote{Another aspect of the qualitative deviation of the intrinsic vortex filament motion in LIA from the actual vortex filament motion arises from the curvature conservation condition concocted by LIA (Svistunov \cite{Svi}).}.

Liquid $^4$He presents a better system than ordinary fluids for the application of LIA, because,
\begin{itemize}
  \item the cores of vortices in superfluid $^4$He are only few angstroms in diameter;
  \item interactions between different segments of a vortex filament are negligible.
\end{itemize}

Taking into account the frictional force exerted by the normal fluid (or quasi-particles) on a scattering vortex line, the self-advection velocity of the vortex line according to the LIA is given by the Hall-Vinen-Bekarevich-Khalatnikov (HVBK) equation (\cite{Hal}, \cite{Hal2} and \cite{Bek})
\be\tag{1}
{\bf v} = \gamma \kappa {\that} \times {\nhat} + \alpha {\that} \times \lp {\bf U} - \gamma \kappa {\that} \times {\nhat} \rp - \alpha^\pr {\that} \times  \lb {\that} \times \lp {\bf U} - \gamma \kappa {\that} \times {\nhat} \rp \rb.
\ee
Here, {\bf U} is the normal fluid velocity (or drift velocity of quasi-particles) taken to be constant in space and time and prescribed\footnote{So this formulation is kinematical in nature - the effect of the vortices on the normal fluid is ignored.} (\cite{Sch}, \cite{Sch2}), $\kappa$ is the average curvature, and ${\that}$ and ${\nhat}$ are unit tangent and unit normal vectors, respectively, to the vortex filament, and $\gamma \equiv \Gamma ln (c/\kappa a_0)$, where $\Gamma$ is the quantum of circulation, $c$ is a constant of order 1 and $a_0 \approx 1 \cdot 3 \times 10^{-8}$ cm is the effective core radius of the filament. $\alpha$ and $\alpha^\pr$ are the friction coefficients which are small (except near the lambda point) so the effect of the friction on the short-term vortex motion appears to be weak. However, the friction term associated with $\alpha$ reflects the fact that a thermal excitation can exchange momentum only in a direction perpendicular to the scattering vortex line and plays the dual roles of driving force and drag force (\cite{Sch}, \cite{Sch2}). One may therefore expect important qualitative effects due to both growth and decay of the vortex line length produced by this friction term. The friction term associated with $\alpha^\pr$ is usually dropped because of the fact that $\alpha > \alpha^\pr$\footnote{Actual numerical values of $\alpha$ and $\alpha^\pr$ at some typical temperatures given in table 1 of Schwarz \cite{Sch} are -
\begin{center}
$T(K) = 1.0 : \alpha = 0.005, ~\alpha^\pr = 0.003$\\
$T(K) = 1.5 : \alpha = 0.073, ~\alpha^\pr = 0.018$.
\end{center}} (Vinen and Nimela \cite{Vin3}). In the absence of the friction force, (1) shows that the vortex filament moves with the {\it local} superfluid velocity. 

Notwithstanding the small values of $\alpha$ and $\alpha^\pr$, the determination of the vortex motion from the HVBK equation (1) is a formidable task and numerical simulations (\cite{Sch}, \cite{Sch2}) have essentially ruled the day. In this paper, we will first do the Hasimoto \cite{Has} type analytical formulation of this problem in the intrinsic geometric parameter space to obtain insights into the underlying physical features as well as highlight difficulties involved in the Hasimoto \cite{Has} type analytical formulation route. We will then give a reformulation of the HVBK equation (1) in the extrinsic vortex filament coordinate space which is known to provide a useful alternative analytical approach in this regard (Shivamoggi \cite{Shi}) - it provides an insight not only into the fundamental importance of the friction term associated with $\alpha$, but also into the previous neglect of the friction term associated with $\alpha^\pr$ (as in the numerical simulations (\cite{Sch}, \cite{Sch2})). We will generalize the formulation in \cite{Shi} and incorporate normal fluid velocity components both {\it along} and {\it transverse} to the vortex filament and investigate their effect on the vortex motion.

\pagebreak

\noindent\Large\textbf{2. Formulation of the HVBK Equation in the Intrinsic Geometric Parameter Space}\\

\large Let us first note that equation (1) can be rewritten as
\be\tag{2}
{\bf v} = \lp 1 - \alpha^\pr \rp \gamma \kappa {\bhat} + \alpha {\that} \times {\bf U} + \alpha \gamma \kappa {\nhat} - \alpha^\pr \lp {\that} \cdot {\bf U} \rp {\that} + \alpha^\pr {\bf U}
\ee
where ${\bhat}$ is the unit binormal vector to the vortex filament. Observe that the friction part associated with $\alpha^\pr$ in the first term on the right hand side can be eliminated by renormalizing the vortex strength $\gamma$,\footnote{This result is also apparent in the vortex-motion formulation via the Gross-Pitaevskii theory (Pismen \cite{Pis}).} albeit an $O (\alpha \alpha^\pr)$ correction to the third term on the right hand side which is small (compared with $O (\alpha)$).

We take the normal fluid velocity to have components both {\it along} and {\it transverse} to the vortex filament,
\be\tag{3}
{\bf U} = U_1 {\that} + U_2 {\nhat} + U_3 {\bhat}.
\ee
Equation (2) then becomes
\be\tag{4}
{\bf v} = \gamma \kappa {\bhat} + \alpha \lp -U_3 {\nhat} + U_2 {\bhat} \rp + \alpha \gamma \kappa {\nhat} + \alpha^\pr \lp U_2 {\nhat} + U_3 {\bhat} \rp.
\ee

Note the Frenet-Serret formulae for the underlying differential geometry,
\be\tag{5}
{\bf x^\pr} = {\that}, ~{\that}^\pr = \kappa {\nhat}, ~{\nhat}^\pr = \tau {\bhat} - \kappa {\that}, ~{\bhat}^\pr = -\tau {\nhat}
\ee
where $\tau$ is the torsion and primes denote differentiation with respect to the arc length s. (5) leads to
\be\notag
\lp {\nhat} + i {\bhat} \rp^\pr = -i \tau \lp {\nhat} + i {\bhat} \rp - \kappa {\that}
\ee
which suggests we introduce (\cite{Has})
\be\tag{6}
{\bf N} \equiv \lp {\nhat} + i {\bhat} \rp e^{i \int \tau (s) ds}, ~\psi \equiv \kappa (s) e^{i \int \tau (s) ds}.
\ee

Note the following relations,
\be\tag{7}
{\bf N} \cdot {\that} = 0, ~{\bf N} \cdot {\bf N} = 0, ~{\bf N} \cdot \bar{\bf N} = 2
\ee
where the bar overhead denotes the complex conjugate of the quantity in question.

(6) leads to
\be\tag{8}
{\bf N}^\pr = -\psi {\that}
\ee
and
\be\tag{9}
{\that}^\pr = Re \lp \psi \bar{\bf N} \rp = \frac{1}{2} \lp \psi \bar{\bf N} + \bar{\psi} {\bf N} \rp.
\ee

On the other hand, we have from equations (4), (5) and (6),
\be\tag{10}
\begin{aligned}
{\bf\dot{\hat{t}}} &= \gamma \lp \kappa^\pr {\bhat} - \kappa \tau {\nhat} \rp + \alpha \gamma \lp \kappa \tau {\bhat} + \kappa^\pr {\nhat} \rp - \alpha \tau \lp U_2 {\nhat} + U_3 {\bhat} \rp\\
&+ \alpha \kappa U_3 {\that} - \alpha \gamma \kappa^2 {\that} + \alpha^\pr \lb U_2 \lp \tau {\bhat} - \kappa {\that} \rp - U_3 \tau {\nhat} \rb\\
&= \gamma Re \lp i \psi^\pr \bar{\bf N} \rp + \alpha \gamma Re \lp \psi^\pr \bar{\bf N} \rp - \alpha \gamma |\psi|^2 {\that} + \alpha Re \lp i \psi \bar{V} \rp {\that} + Re \lb \lp i \alpha + \alpha^\pr \rp V^\pr \bar{\bf N} \rb\\
&= i \frac{\gamma}{2} \lp \psi^\pr \bar{\bf N} - \bar{\psi}^\pr {\bf N} \rp + \frac{\alpha \gamma}{2} \lp \psi^\pr \bar{\bf N} + \bar{\psi}^\pr {\bf N} \rp - \alpha \gamma |\psi|^2 {\that} + i \frac{\alpha}{2} \lp \psi \bar{V} - \bar{\psi} V \rp {\that}\\
&+ i \frac{\alpha}{2} \lp V^\pr \bar{\bf N} - \bar{V}^\pr {\bf N} \rp + \frac{\alpha^\pr}{2} \lp V^\pr \bar{\bf N} + \bar{V}^\pr {\bf N} \rp
\end{aligned}
\ee
where the dot overhead denotes differentiation with respect to time $t$, and
\be\tag{11}
V \equiv \lp U_2 + i U_3 \rp e^{i \int \tau (s) ds}.
\ee

Let
\be\tag{12}
\dot{\bf N} = \sigma {\bf N} + \beta {\that}
\ee
and on using (7) and (10),
\be\tag{13}
\sigma + \bar{\sigma} = \frac{1}{2} \lp \dot{\bf N} \cdot \bar{\bf N} + \dot{\bar{\bf N}} \cdot {\bf N} \rp = \frac{1}{2} \frac{\pa}{\pa t} \lp {\bf N} \cdot \bar{\bf N} \rp = 0 ~~\text{or} ~~\sigma = iR
\ee
and
\be\tag{14}
\beta = \dot{\bf N} \cdot {\that} = -{\bf N} \cdot \dot{{\that}} = -\lp i \gamma + \alpha \gamma \rp \psi^\pr - i \alpha V^\pr
\ee
where R is some real-valued function.

Substituting (13) and (14), (12) becomes
\be\tag{15}
\dot{\bf N} = iR {\bf N} + \lb -\lp i + \alpha \rp \gamma \psi^\pr - i \alpha V^\pr \rb {\that}.
\ee

Taking the time derivative of equation (8) and the s-derivative of equation (15), and using equations (8) - (10), we obtain
\be\tag{16}
\begin{aligned}
\dot{\bf N}^\pr = &-\psi {\bf\dot{{\that}}} - \dot{\psi} {\that} = -\psi \lb i \frac{\gamma}{2} \lp \psi^\pr \bar{\bf N} - \bar{\psi}^\pr {\bf N} \rp + \frac{\alpha \gamma}{2} \lp \psi^\pr \bar{\bf N} + \bar{\psi}^\pr {\bf N} \rp - \alpha \gamma |\psi|^2 {\that} \right.\\
&\left. + i \frac{\alpha}{2} \lp \psi \bar{V} - \bar{\psi} V \rp {\that} + i \frac{\alpha}{2} \lp V^\pr \bar{\bf N} - \bar{V}^\pr {\bf N} \rp \rb - \dot{\psi} {\that} + \frac{\alpha^\pr}{2} \lp V^\pr \bar{\bf N} + \bar{V}^\pr {\bf N} \rp
\end{aligned}
\ee
and
\be\tag{17}
\dot{\bf N}^\pr = i \lp R^\pr {\bf N} - R \psi {\that} \rp - \lb \lp i \gamma + \alpha \gamma \rp \psi^{\pr \pr} + i \alpha V^{\pr \pr} \rb {\that} - \lb \lp i \gamma + \alpha \gamma \rp \psi^\pr + i\alpha V^\pr \rb \frac{1}{2} \lp \psi \bar{\bf N} + \bar{\psi} {\bf N} \rp.
\ee

Equating (16) and (17), we obtain the following coupled equations,
\be\tag{18}
-\dot{\psi} + \alpha \gamma |\psi|^2 \psi - i \frac{\alpha}{2} \psi \lp \psi \bar{V} - \bar{\psi} V \rp = -iR \psi - \lb \lp i + \alpha \rp \gamma \psi^{\pr \pr} + i \alpha V^{\pr \pr} \rb
\ee
\be\tag{19}
i \frac{\gamma}{2} \psi \bar{\psi}^\pr - \frac{\alpha \gamma}{2} \psi \bar{\psi}^\pr + i \frac{\alpha}{2} \psi \bar{V}^\pr - \frac{\alpha^\pr}{2} \psi \bar{V}^\pr = iR^\pr - \frac{1}{2} \lb \lp i + \alpha \rp \gamma \psi^\pr + i \alpha V^\pr \rb \bar{\psi}.
\ee

The almost impossibility of decoupling equations (18) and (19), as they are, highlights the enormous difficulty in fully determining the vortex motion in an analytic way from the HVBK equation (1). In order to make further progress, appropriate approximations are necessary. Recognizing that $\alpha$ and $\alpha^\pr$ are very small, in a first approximation, equations (18) and (19) may be decoupled to give the nonlinear Schrodinger equation,
\be\tag{20}
\frac{1}{i} \dot{\Phi} \approx \frac{\gamma}{2} |\Phi|^2 \Phi + \gamma \Phi^{\pr \pr}
\ee
where,
\be\tag{21}
\Phi \equiv \psi + \frac{\alpha}{\gamma} V.
\ee
Equation (20) is the same as that for the ordinary fluid case (\cite{Has}) with the superfluid effects now represented by a renormalization of the ``wave function" $\psi$, as in (21), in the first approximation.

Noting (6) and (11), (21) may be approximated as follows,
\be\tag{22}
\begin{aligned}
\Phi &= \kappa \lb 1 + \frac{\alpha}{\gamma \kappa} \lp U_2 + iU_3 \rp \rb e^{i \int \tau ds} \approx \kappa e^{\frac{\alpha}{\gamma \kappa} \lp U_2 + i U_3 \rp + i \int \tau ds}\\
& = \lp \kappa e^{\frac{\alpha}{\gamma \kappa} U_2} \rp e^{i \int \lb \tau + \frac{\alpha}{\gamma} \lp \frac{U_3}{\kappa} \rp^\pr \rb ds} \approx \lp \kappa + \frac{\alpha}{\gamma} U_2 \rp e^{i \int \lb \tau + \frac{\alpha}{\gamma} \lp \frac{U_3}{\kappa} \rp^\pr \rb ds}.
\end{aligned}
\ee
(22) implies that, to first approximation, the friction term with $\alpha$ via
\begin{itemize}
  \item the normal fluid velocity component along ${\nhat}$ serves to modify the curvature $\kappa$;
  \item the normal fluid velocity component along ${\bhat}$ serves to modify the torsion $\tau$;
\end{itemize}
as to be expected.

In recognition of the enormous difficulty involved in making further progress in dealing with equations (18) and (19) in an analytic way, we now consider a reformulation of the HVBK equation (1) in the extrinsic vortex filament coordinate space.

\vspace{.3in}

\noindent\Large\textbf{3. Reformulation of HVBK Equation in the Extrinsic Vortex Filament Coordinate Space: Generalized Normal Fluid Flow}\\

\large In the extrinsic vortex filament coordinate space formulation we consider only small-amplitude vortex motions in the LIA model. So this approach circumvents the difficulties hampering the large-amplitude solutions due to violation of the basic premise of LIA that distant parts of the vortex filament remain sufficiently separated during the motion.\footnote{This is like avoiding the wave breaking situation during a wave steepening process in nonlinear hyperbolic systems (Kuznetsov and Ruban \cite{Kuz}).}

Consider the vortex filament essentially aligned along the x-axis (Dmitreyev \cite{Dmi}, Shivamoggi and van Heijst \cite{Shi2}). In the following development we generalize the formulation in \cite{Shi} and incorporate the normal fluid velocity components both {\it along} and {\it transverse} to the vortex filament, so we take ${\bf U} = U_1 {\ihat}_x + U_2 {\ihat}_y + U_3 {\ihat}_z$. Equation (1) then becomes
\be\tag{23}
{\bf v} = \lp 1 - \alpha^\pr \rp \gamma \kappa {\that} \times {\nhat} + \alpha {\that} \times {\bf U} + \alpha \gamma \kappa {\nhat} - \alpha^\pr U_1 {\that} + \alpha^\pr {\bf U}.
\ee
Since we are assuming small-amplitude vortex motions, we assume the deviations of the vortex filament from the x-axis to be small, so we have
\be\tag{24}
{\bf r} = x {\ihat}_x + y (x, t) {\ihat}_y + z (x, t) {\ihat}_z.
\ee

We then have
\be\tag{25}
{\bf v} \equiv \frac{d {\bf r}}{dt} = y_t {\ihat}_y + z_t {\ihat}_z
\ee
\be\tag{26}
{\that} \equiv \frac{d {\bf r}}{ds} = \frac{d {\bf r}}{dx} \frac{dx}{ds} = \lp {\ihat}_x + y_x {\ihat}_y + z_x {\ihat}_z \rp \frac{dx}{ds}
\ee
\be\tag{27}
\begin{aligned}
\kappa {\nhat} \equiv \frac{d {\that}}{ds} = \frac{d {\that}}{dx} \frac{dx}{ds} \approx &-\lp y_x y_{xx} + z_x z_{xx} \rp \frac{dx}{ds} ~{\ihat}_x + \lb y_{xx} \frac{dx}{ds} - \lp y^2_x y_{xx} + y_x z_x z_{xx}\rp \rb \frac{dx}{ds} ~{\ihat}_y\\
& + \lb z_{xx} \frac{dx}{ds} - \lp z_x y_x y_{xx} + z^2_x z_{xx} \rp \rb \frac{dx}{ds} ~{\ihat}_z.
\end{aligned}
\ee
where,
\be\tag{28}
\frac{dx}{ds} = \lp 1 + y^2_x + z^2_x \rp^{-1/2} \approx 1 - \frac{1}{2} \lp y^2_x + z^2_x \rp.
\ee

Substituting (25) - (28), equation (23) gives
\be\tag{29}
y_t = -\lp 1 - \alpha^\pr \rp \gamma z_{xx} + \frac{3 \gamma}{2} \lp y^2_x + z^2_x \rp z_{xx} + \alpha U_1 z_x - \alpha U_3 + \alpha \gamma y_{xx} - \alpha^\pr U_1 y_x + \alpha^\pr U_2
\ee
\be\tag{30}
z_t = \lp 1 - \alpha^\pr \rp \gamma y_{xx} - \frac{3 \gamma}{2} \lp y^2_x + z^2_x \rp y_{xx} - \alpha U_1 y_x + \alpha U_2 + \alpha \gamma z_{xx} - \alpha^\pr U_1 z_x + \alpha^\pr U_3.
\ee

Putting,
\be\tag{31}
\Phi \equiv y + iz, ~V \equiv U_2 + i U_3
\ee
and keeping only linear terms associated with $\alpha$ and $\alpha^\pr$ (since $\alpha$ and $\alpha^\pr$ are very small), equations (29) and (30) can be combined to give the nonlinear Schrodinger equation
\be\tag{32}
\frac{1}{i} \lb \Phi_t - \lp i \alpha + \alpha^\pr \rp V + \alpha^\pr U_1 \Phi_x \rb + \alpha U_1 \Phi_x = \lp 1 - \alpha^\pr \rp \gamma \Phi_{xx} - i \alpha \gamma \Phi_{xx} - \frac{3 \gamma}{2} |\Phi^2_x| \Phi_{xx}.
\ee
Equation (32) describes the propagation of weakly nonlinear Kelvin waves on a vortex filament in a superfluid.

Equation (32) admits a solution of the form,
\be\tag{33}
\Phi (x, t) = \chi (x, t) + \lp i \alpha + \alpha^\pr \rp V t
\ee
where $\chi (x, t)$ satisfies the equation given in \cite{Shi} corresponding to no normal fluid flow velocity {\it transverse} to the vortex filament,
\be\tag{34}
\frac{1}{i} \lp \chi_t + \alpha^\pr U_1 \chi_x \rp + \alpha U_1 \chi_x = \lp 1 - \alpha^\pr \rp \gamma \chi_{xx} - i \alpha \gamma \chi_{xx} - \frac{3 \gamma}{2} |\chi_x|^2 \chi_{xx}.
\ee
(33) shows that the normal fluid velocity components {\it transverse} to the vortex filament via the friction terms associated with $\alpha$ and $\alpha^\pr$ merely produce an algebraic growth of the Kelvin waves on the vortex filament, as per
\be\tag{35}
\left.
\begin{matrix}
y \sim \lp \alpha^\pr U_2 - \alpha U_3 \rp t\\
z \sim \lp \alpha U_2 + \alpha^\pr U_3 \rp t.
\end{matrix}
\rc
\ee
The friction term associated with $\alpha^\pr$ on the left hand side in equation (34) can be eliminated by introducing the Galilean transformation
\be\tag{36}
q (x, t) \Rightarrow q (\xi, t), ~\xi \equiv x - \alpha^\pr U_1 t.
\ee
The friction term associated with $\alpha^\pr$ in the first term on the right hand side can again be eliminated by renormalizing the vortex strength $\gamma$, as clarified in detail in \cite{Shi}.

\vspace{.3in}

\noindent\Large\textbf{4. Vortex Motion in an Ordinary Fluid}\\

\large In an ordinary fluid, the friction terms associated with $\alpha$ and $\alpha^\pr$ disappear, and equation (34) becomes
\be\tag{37}
\frac{1}{i} \chi_t = \gamma \lp 1 - \frac{3}{2} |\chi_x|^2 \rp \chi_{xx}.
\ee

Letting $\chi \sim e^{ikx}$ in the nonlinear dispersion term, equation (36) may be approximated by
\be\tag{38}
\frac{i}{\gamma} \chi_t = \lp -\frac{\pa^2}{\pa x^2} - \frac{3}{2} k^4 |\chi|^2 \rp \chi.
\ee
Equation (38) may be viewed as a Schrodinger equation for quasi-particles which are trapped in a self-generated attractive potential well. The trapping process deepens the potential well (the depth being proportional to the particle density $|\chi|^2$) which enhances trapping further. The end result of this runaway process is the development of solitary waves (Whitham \cite{Whi}). This scenario continues to hold even in superfluid $^4$He, as we will see below, provided the friction coefficients $\alpha$ and $\alpha^\pr$ remain small.

\vspace{.3in}

\noindent\Large\textbf{5. Nonlinear Localized Structures on a Vortex Filament}\\

\large Let us look for a nonlinear localized Kelvin stationary wave solution \cite{Shi},
\be\tag{39}
\chi (x, t) = \nu \psi \lp \xi - u \gamma t \rp e^{i \lp \sigma \xi - \tilde{c} \gamma t \rp + \mu t}
\ee
and choose the parameters as follows,
\be\tag{40}
\sigma = \frac{u}{2}, ~\beta \equiv \sigma^2 - \tilde{c}, ~\mu = \frac{\alpha u}{2} \lp U_1 - \frac{u}{2}\rp.
\ee
The first relation in (40) implies that the velocity of propagation of this nonlinear structure is twice the torsion, while $\beta$ is a measure of the curvature. 

Using (39) and (40), and assuming $\psi$ is slowly-varying, equation (34) leads to \cite{Shi}
\be\tag{41}
\psi^{\pr \pr} - \beta \psi + \frac{3 \nu \sigma}{2}^4 \psi^3 = 0
\ee
with an envelope solitary-wave solution
\be\tag{42}
\psi = \sqrt{\frac{4 \beta}{3 \nu \sigma^4}} ~\text{sech} \sqrt{\beta} \lp \xi - 2 \sigma \gamma t \rp
\ee
describing a propagating damped or growing Kelvin wave on a vortex filament in superfluid $^4$He. The vortex kink growth associated with the friction coefficient $\alpha$, implied by (39) and (40), as mentioned in \cite{Shi}, has qualitative similarities with the Donnelly-Glaberson instability (Cheng et al. \cite{Che}, Glaberson et al. \cite{Gla}) of Kelvin waves on a vortex driven by the normal fluid velocity component {\it along} the vortex filament. On the other hand, it should be noted that the friction term associated with $\alpha^\pr$, even in the presence of the normal fluid velocity components {\it transverse} to the vortex filament, as per (33) and (35), merely produces an algebraic growth of the Kelvin waves hence providing further justification for the neglect of this term.

In the absence of the normal fluid flow velocity components {\it transverse} to the vortex filament ($V = 0$), for the case $u = 2 U_1$, the vortex kink is undamped (the nonlinearity then balancing both the dispersion and the mutual friction) and the vortex kink amplitude as well as its propagation speed are totally determined by the normal fluid velocity $U_1$. So, this case corresponds to a {\it one-parameter} family (interestingly befitting the quantum mechanical regime) of envelope solitary wave solutions while the usual nonlinear Schrodinger envelope solitary waves are characterized by two parameters. 

Thus, the friction term associated with $\alpha$, unlike that associated with $\alpha^\pr$, causes qualitative changes in the vortex kink dynamics characteristics\footnote{The quantitative effects of $\alpha$ may become non-negligible, however, if the normal-fluid velocity $U_1$ is very high.}. The rotating planar vortex filament problem in superfluid $^4$He sheds further light on this aspect, as seen below.

\vspace{.3in}

\noindent\Large\textbf{6. Parametric Amplification of Kelvin Waves}\\

\large Consider a planar vortex filament given by $y = y(x)$ which is lying in the x, y-plane and is rotated with a uniform angular velocity $\Omega$ about the x-axis (this problem was considered by Hasimoto \cite{Has2} in the ordinary fluid case). Noting in this case,
\be\tag{43}
{\that} = < 1, y_x, 0 > \frac{1}{\sqrt{1 + y^2_x}}
\ee
\be\tag{44}
\kappa {\nhat} \equiv \frac{d {\that}}{ds} = \frac{d {\that}}{dx} \frac{dx}{ds} = \la -\frac{y_x y_{xx}}{\lp 1 + y^2_x \rp^{3/2}}, ~\frac{y_{xx}}{\lp 1 + y^2_x \rp^{3/2}}, ~0 \ra \frac{1}{\sqrt{1 + y^2_x}}
\ee
taking ${\bf U} = U_1 {\ihat}_x$, and putting
\be\tag{45}
y_x = \tan \theta
\ee
the z-component of equation (23) leads to \cite{Shi},
\be\tag{46}
\lp 1 - \alpha^\pr \rp \gamma \frac{d^2 \theta}{ds^2} - \alpha U_1 \cos \theta \cdot \frac{d \theta}{ds} + \Omega \sin \theta = 0.
\ee
The friction term associated with $\alpha^\pr$ in equation (46) can be eliminated by renormalizing the vortex strength $\gamma$, as in Section 3.

In the ordinary-fluid limit, equation (46) constitutes {\it Euler's elastica} (Love \cite{Lov}), i.e., the finite deformation of a plane elastic filament of flexural rigidity B under the action of the thrust F applied at its ends,
\be\tag{47}
B \frac{d^2 \theta}{ds^2} + F \sin \theta = 0.
\ee

On the other hand, assuming $| \Omega / \gamma | \ll 1$ and expanding in powers of $\theta$, equation (46) leads to the Van der Pol equation \cite{Shi},
\be\tag{48}
\gamma \frac{d^2 \theta}{ds^2} + \alpha U_1 \lp \frac{\theta^2}{2} - 1 \rp \frac{d \theta}{ds} + \Omega \theta = 0
\ee
which reveals that there is decay/growth of the vortex line length if $|\theta|^2 \gtrless 4$ (and the normal fluid velocity is in the same direction as that of vorticity in the undisturbed vortex filament) via a normal fluid flow driven instability\footnote{Experiments by Sato et al. \cite{Sat} on vortex dynamics in $^4$He near the lambda point indicated vortex growth due to an instability.}, as seen in Section 5.

Equation (46) also indicates the possibility of parametric amplification of Kelvin waves by the friction force. In order to see this more clearly, note that an approximate solution of equation (46) is
\be\tag{49a}
\theta \approx \sin \omega s + O (\alpha), ~\omega^2 \equiv \Omega/\gamma
\ee
which, for $\omega s \ll 1$, may be approximated further by
\be\tag{49b}
\theta \approx \omega s.
\ee
Substituting (49b) in the friction term of equation (46), and linearizing the third term, we obtain
\be\tag{50}
\frac{d^2 \theta}{ds} - \lp \frac{\alpha U_1}{\gamma} \rp \cos \omega s \cdot \frac{d \theta}{ds} + \omega^2 \theta = 0.
\ee
On introducing the Liouville transformation,
\be\tag{51}
\theta (s) = u (s) e^{\displaystyle \lp \frac{\alpha U_1}{2 \gamma \omega} \rp \sin \omega s}
\ee
equation (50) becomes
\be\tag{52}
\frac{d^2 u}{d s^2} + \lb \omega^2 - \lp \frac{\alpha U_1 \omega}{2 \gamma} \rp \sin \omega s \rb u = 0
\ee
which is Mathieu's equation describing the parametric amplification of Kelvin waves by the friction force. Equation (52) reveals again that the friction term associated with $\alpha$ influences the vortex motion in a qualitative way.

\vspace{.3in}

\noindent\Large\textbf{7. Discussion}\\

\large It is now generally recognized that the phenomenological model of quantized vortices as classical vortex filaments subject to an effective frictional force (simulating interactions with thermal excitations in superfluid $^4$He) provides a useful approach to investigate vortex dynamics in superfluid $^4$He. The theoretical formulations developed in this paper provide insight into the fundamental importance of the friction terms associated with the friction coefficient $\alpha$ as well as the previous neglect of the friction term associated with the friction coefficient $\alpha^\pr$. The friction term associated with $\alpha$ via the normal fluid velocity component {\it along} the vortex filament drives the Kelvin waves unstable. On the other hand, the friction term associated with $\alpha^\pr$, even in the presence of the normal fluid velocity components {\it transverse} to the vortex filament, is shown to produce merely an algebraic growth of the Kelvin waves hence providing further justification for the neglect of this term. The instability produced by the friction term associated with $\alpha$ is shown to be of parametric amplification type in certain cases. This term, unlike that associated with $\alpha^\pr$, therefore causes qualitative changes in the vortex kink dynamics characteristics.

\vspace{.3in}

\noindent\Large\textbf{Acknowledgments}\\

\large This work was carried out when the author held a visiting research appointment at the Eindhoven University of Technology supported by a grant from The Netherlands Organization for Scientific Research (NWO). The author is thankful to Professor Gert Jan van Heijst for his hospitality and helpful discussions. The author is thankful to Professors K. R. Sreenivasan, Dan Lathrop and Drs. Yuki Sato and Demosthenes Kivotides for helpful remarks and Professor Michael Johnson for helpful discussions.

\end{document}